# Electrical Oscillation in Pt/VO$_2$ Bilayer Strips

Ying Wang,[1] Jianwei Chai,[2] Shijie Wang,[2] Long Qi,[1] Yumeng Yang,[1,2] Yanjun Xu,[1] Hidekazu Tanaka,[3] and Yihong Wu[1,a]

[1] *Department of Electrical and Computer Engineering, National University of Singapore, 4 Engineering Drive 3, Singapore 117583*

[2] *Institute of Materials Research & Engineering, Agency for Science, Technologies and Research, 3 Research Link, Singapore 117602*

[3] *Institute of Scientific and Industrial Research, Osaka University, 8-1 Mihogaoka, Ibaraki, Osaka 567-0047, Japan*

We report on the observation of stable electrical oscillation in Pt/VO$_2$ bilayer strips, in which the Pt overlayer serves the dual purposes of heating up the VO$_2$ and weakening the electric field in the VO$_2$ layer. Systematic measurements in an ultrahigh vacuum nanoprobe system show that the oscillation frequency increases with the bias current and/or with decreasing device dimension. In contrast to most VO$_2$-based oscillators reported to date, which are electrically triggered, current-induced Joule heating in the Pt overlayer is found to play a dominant role in the generation of oscillation in Pt/VO$_2$ bilayers. A simple model involving thermally triggered transition of VO$_2$ on a heat sink is able to account for the experimental observations. The results in this work provide an alternative view of the triggering mechanism in VO$_2$-based oscillators.

[a] Author to whom correspondence should be addressed: elewuyh@nus.edu.sg

**I. INTRODUCTION**

Vanadium dioxide (VO$_2$) has attracted much interest due to its ultrafast first-order insulator-to-metal transition near room temperature (~341 K) and outstanding thermodynamic stability.[1-3] It has been reported that the transition of VO$_2$ can be triggered by different external stimuli such as thermal,[4] electrical,[5-7] optical[8] and strain[9] excitations, and that it is accompanied by a structural transition from a monoclinic insulating phase (at low temperature) to a tetragonal metallic phase (at high temperature).[10] Despite having been investigated for a few decades, the physics underlying VO$_2$ transition is still under debate. While some studies favor a Peierls transition picture,[11-13] most other evidence suggests a Mott transition model.[14-17] From application point of view, the abrupt transition with a very large change in resistivity up to over four orders of magnitude opens up new opportunities for a variety of potential solid-state applications such as, but not limited to, electrical oscillators, gated field effect switches, memristors, thermal sensors and chemical sensors.[18] Among them VO$_2$-based oscillator appears to be particularly interesting due to its simplicity in implementation and the ease of frequency modulation as compared to conventional oscillators which usually consist of active devices, piezoelectric components and/or RC circuits.

Most of the VO$_2$-based oscillators reported to date typically consist of a simple two-terminal VO$_2$ device (either in-plane or out-of-plane), a serial resistor typically in the kΩ range (either externally connected or from the measuring circuitry), and a voltage source (a constant bias voltage with/without a superimposed pulse voltage).[19-27] When the applied voltage exceeds a certain threshold, the circuit will oscillate with a frequency up to sub-MHz. It has also been shown that further optimization of the external circuit components (*i.e.* resistor or capacitor), down scaling of the size of VO$_2$ pattern, doping with tungsten or illumination with an IR laser can help increase the maximum frequency to the MHz range.[23,26,28,29] The underlying mechanism of oscillation has often been attributed to the alternative occurrence of insulator-to-metal transition (IMT) and metal-to-insulator transition (MIT) of VO$_2$, which causes alternative division of the bias voltage between the VO$_2$ device and the serial resistor. A distinct feature of the oscillation waveform is that the voltage across (and the electric field in) the VO$_2$ device drops abruptly upon reaching a certain critical value (0.71 – 65 V/μm in terms of electric field) regardless of the bias method. Sakai found that the oscillation only occurs for certain combination of the source voltage and load resistance, and suggested an electric-field-induced transition model.[21] Kim's group proposed a modified percolative-avalanche model to account for the observed oscillation and also identified the electric field as the most likely candidate for driving the $VO_2$ transition.[22,23,30] Beaumont *et al.* studied the oscillation behavior of out-of-plane devices connected in series with a small resistor and biased with a DC current source, and suggested that the fast transition time (10 – 12 ns) of the device cannot be explained by a simple Joule heating model alone.[24] On the other hand, a Joule-heating-induced transition picture has also been supported by some works. Fisher observed oscillations in voltage-biased $VO_2$ needles in series with 42 kΩ resistors and associated it with travelling semiconductor domains driven by Peltier effect at their boundary.[20] This explanation is adopted by Gu *et al.* to account for the experimental results of their recent work in which W-doped $VO_2$ nanobeams are connected to a parallel shunt capacitor (~100 pF) and biased with a current source.[28] The authors conclude that the oscillation is dictated by the Joule-heating-induced MIT, heat dissipation, the Peltier effect and the axial drift of single metal-insulator domain walls. Given the fact that the Joule heating effect on the transition of $VO_2$ is closely related to the device geometry, the contact material and the bias current,[31] current understanding on the role of Joule heating in $VO_2$-based oscillators is still elusive. To shed further light on the oscillation triggering mechanism, we devised a structure which consists of only a Pt/$VO_2$ bilayer and studied its oscillation characteristics under a constant bias current. In this design, since the current passes mainly through the Pt layer when $VO_2$ is in the insulating state, the role of electric field in triggering the oscillation should be reduced. Instead, Joule heating by the current in the Pt layer is anticipated to play a dominant role. Therefore, the bilayer device configuration will help to provide an alternative view of the triggering mechanism in $VO_2$-based oscillators. The remaining of this paper is organized as follows. Device fabrication, experimental setup and procedure of electrical measurements are described in Section II. Section III begins with presenting the experimentally observed dependence of oscillation on the device dimension and bias current, and then introduces a simple model based on thermally triggered transition of the $VO_2$ to account for the experimental results. The last part of Section III briefly discusses the potential role of electric field in triggering $VO_2$ transition in the investigated devices. Section IV provides a summary of the work.

**II. EXPERIMENTAL**

VO$_2$ (001) films with a nominal thickness of 100 nm are grown on c-cut sapphire substrates at 600°C by DC unbalanced magnetron sputtering of a vanadium target (99.9%) with a mixture of Ar:O$_2$ gas (at a ratio of 6:1) and at a power of 275 W. The base pressure of the system is $7\times10^{-7}$ mbar. Four-probe electrical measurements on the as-grown thin film show an abrupt resistance change of almost four orders of magnitude (i.e. from 10000 to 2 Ω) upon phase transition at round 345 K, suggesting that the VO$_2$ is of high quality.[32] Thin Pt strips ($t_0$ = ~3 nm) of different widths are subsequently patterned and deposited onto the as-grown VO$_2$ film using a laser writer (Microtech LW405) for creating the resist pattern and RF sputtering for deposition of thin Pt layer. These Pt strips serve as the local heating elements. A schematic of the Pt/VO$_2$ devices is shown in Fig. 1(a).

All electrical measurements are performed in an ultrahigh vacuum (UHV) Omicron nanoprobe system to ensure good reproducibility and reliability of the experimental results. An *in-situ* scanning electron microscope (SEM) and two independent piezo-electrically driven nanoprobes with auto-approaching capability allow for precise control of the probe positions and real-time monitoring of the sample surface.[33] The procedure of measurements is as follows. Two sharp tungsten probes are first approached to and firmly pressed onto the two ends of a long 1μm-wide Pt/VO$_2$ strip (hereafter denoted as Type A device) to form low-resistance contacts. A small DC bias current ($I_0$) is then applied to the probes with a Keithley 6221 source meter. $I_0$ is subsequently ramped up slowly, during which the voltage across the Pt/VO$_2$ device is recorded by a digital oscilloscope. The dependence of oscillation on the Pt/VO$_2$ channel length is readily studied by stepping one probe towards the other on the strip, as illustrated in Fig. 1(a) and (c). In order to investigate the dependence of oscillation on the strip width (W), a second type of Pt/VO$_2$ device (Type B) with different W (1 – 8 μm) and a fixed length (L) of 40 μm has also been fabricated and measured [Fig. 1(b)]. The electron beam from the SEM is normally blanked off during the measurements. All the measurements were carried out at room temperature.

### III. RESULTS AND DISCUSSION

**A. Dependence of oscillation on the device dimensions and the bias current**

The inset of Fig. 2(a) shows the typical oscillation obtained by passing a DC current of 5.6 mA through a device with a dimension of 40 μm (L) × 2 μm (W) in a time span of 150 μs. The oscillation frequency is ~99 kHz as shown by the very sharp peak in the corresponding Fourier transform [Fig. 2(a)]. For this particular device, the oscillation kicks in at a critical current of $I_{c1}$ = 2.488 mA as seen in Fig. 2(b), in which all curves except for the lowest one have been vertically shifted for clarity. The oscillation frequency stabilizes within a few μA above $I_{c1}$ and then monotonically increases with increasing the bias current. A closer look at the waveforms in Fig 2(b) reveals that the rising time is shorter at larger current while the falling time remains almost constant. The oscillation disappears when the bias current exceeds another critical bias current ($I_{c2}$). Similar dependence of the oscillation frequency and waveform on the bias current is also observed in Type A device with an L ranging from 1 to 32 μm and a fixed W = 1 μm, which are summarized in a color contour plot shown in Fig. 3(a). In the plot, the lower ($I_{c1}$) and upper ($I_{c2}$) critical currents are indicated as dotted lines. A window for frequency modulation (Region II) is clearly seen in-between two non-oscillating regions (I and III). For the sake of clarity, the relations between the oscillation frequency and the bias current for different L are re-plotted in Fig. 3(b). It is found that the tunable frequency range of longer devices is lower than that of shorter devices (for example, 6.5 - 105 kHz for L = 32 μm and 128 - 262 kHz for L = 1 μm). Besides, the frequency of shorter devices is more sensitive to the change of bias current, enabling more efficient frequency modulation.

An oscillation window is also observed in Type B devices [shaded region in the inset of Fig. 3(c)]. Oscillation is normally not observed in devices with W > 8 μm. The frequency increases monotonically with increasing current, as expected. Interestingly, thinner devices (i.e. smaller W) generate a higher frequency than wider devices at the same bias current. This suggests that the frequency can be further improved by scaling down the dimensions of the oscillator.

**B. Proposed model for the electrical oscillation**

The aforementioned oscillatory behavior can be qualitatively understood as follows. When $VO_2$ is in the insulating state, most of the current applied will flow through the Pt layer. The thin Pt layer is highly resistive with a resistivity of 639 nΩ·m and 1400 nΩ·m for Type A and Type B devices (to be deduced later), respectively. This will lead to Joule heating and temperature rise of the Pt layer. When the temperature in $VO_2$ exceeds the phase transition temperature, its resistance will drop, diverting most of the current into the $VO_2$ layer. This in turn will reduce Joule heating and eventually restores the insulating state of $VO_2$. A question naturally arises here: what determines the oscillation frequency? To facilitate the discussion, we begin with the state that the $VO_2$ is in a metallic state and that the inevitable parasitic capacitance ($C_{para}$) from the circuit is almost completely discharged (hereafter we call it t = 0 state). At this state, the sudden drop of device temperature triggers an MIT inside the $VO_2$ which causes an abrupt increase in the $VO_2$ resistance [Fig. 4(a)]. In the period of $0 < t < t_c$, $C_{para}$ is charged towards a target voltage ($V_{target} = I_0 \times R_{DUT}$) and this is the reason why voltage (V) across the device under test (DUT) gradually increases in this period. Accompanying this process is the increase of the current in the Pt overlayer ($I_J$) and thus the increase of heat generation rate and the device temperature. At $t = t_c$ and $V = V_{max}$, the high device temperature triggers an IMT inside the $VO_2$ and an abrupt decrease in the device resistance. Immediately following the IMT, $C_{para}$ starts to discharge and V decreases rapidly. Since the time constant of the discharging process is mainly determined by the small resistance of $VO_2$, the rate of discharge between $t_c < t < t_2$ will be much faster than the rate of charging in the duration $0 < t < t_c$, which is limited by the much larger $R_{Pt}$. The rate of heat generation in the device is minimal since most of the bias current passes the low resistance $VO_2$. Due to the rapid heat dissipation through the sapphire substrate underneath, the device temperature will decrease until another MIT is triggered at $t = t_2$. Thus, the same sequence of processes described above will be repeated. The charging and discharging time constants determine the oscillation frequency. As seen in Fig. 3(c), narrower Type B devices (thus a larger $R_{DUT}$) oscillate at a higher frequency than the wider ones at the same bias current despite a larger time constant for charging. This is because the current density and thus the rate of heat generation in Pt is larger in narrower devices, resulting in a faster heat accumulation in the device. On the other hand, the current density in the Pt overlayer of all Type A devices is the same due to a fixed W. In this case, time constant is the dominant factor because a larger time constant results in a slower increase of the current flowing through Pt. Thus, longer devices (thus a larger time constant) oscillate at a lower frequency [Fig. 3(a)]. Oscillatory behavior does not occur for $I_0 < I_{c1}$ because the current-induced heating in the Pt overlayer is insufficient to bring the $VO_2$ to the critical IMT temperature. On the other hand, the device temperature is maintained above the critical MIT point for $I_0 > I_{c2}$, and the $VO_2$ is unable to restore to the insulting state.

For a more quantitative understanding of the process, a simple model is proposed in Fig. 4(b) for the Pt/VO$_2$ oscillator, in which the VO$_2$ is treated as a switch (S) in series with a resistor ($R_I$ or $R_M$, depending on the state of VO$_2$). The resistance of metallic (or insulating) VO$_2$ is denoted as $R_M$ (or $R_I$). This assumption is adequate because the transition of VO$_2$ between the metallic and insulating states is very rapid in terms of change in electrical conductivity.[6,34] The MIT (or IMT) of VO$_2$ can thus be seen as simply connecting the switch S to $R_I$ (or $R_M$). The capacitance resulted from metallic nano-islands in VO$_2$ is much smaller than $C_{para}$ and is thus neglected.[23,24] Figures 4(c) and (d) show a snapshot of a single period of oscillation waveform from Type A ($I_0$ = ~1.45 mA) and Type B ($I_0$ = 5 mA) devices, respectively. All curves except for the lowest one have been vertically shifted by an equal amount for the sake of clarity. Although the waveforms of Type A devices with L < 6 μm is not shown in the figures due to their much smaller scale, the discussion below applies to them equally. The experimental charging curves (symbols) have been fitted (solid curves) by the expression below:

$$V(t) = V_{target}(1 - e^{-\frac{t-t_s}{\tau}}), \qquad (1)$$

where the time constant $\tau = R_{DUT} \times C_{para}$, $V_{target} = R_{DUT} \times I_0$ and a shift of $t_s$ that accounts for the small but nonzero voltage at t = 0 are the fitting parameters. It can be seen that the fitting is generally satisfactory, suggesting the adequateness of the assumption. The slight deviation of the fitting curves and the experimental data around t = 0 originates from the finite transition time of VO$_2$ as compared to an ideal switch. Figures 5(a) and (c) show the $\tau$ used for the fittings together with the $R_{Pt}$ calculated from $R_{DUT}$ by using an experimental resistivity of 7.88 Ω·cm for insulating VO$_2$ and with current spreading in the VO$_2$ thin film taken in to account. As expected, $R_{Pt}$ depends linearly on the strip length while it is inversely proportional to the strip width. The Pt resistivity for Type A and B devices can be calculated from the fitting curves in Figs. 5(a) and (c) to be 639 nΩ·m and 1400 nΩ·m, respectively. These values are quite large as compared to bulk Pt resistivity because of the small thickness of the Pt layer. The capacitance of both types of devices has also been extracted by dividing $R_{DUT}$ from $\tau$ and plotted in Fig. 5(b) and (d). Interestingly, $C_{para}$ depends on L but it is almost independent of W. This is presumably resulted from the electrostatic interaction between the two probes. Further systematic investigation is required to reveal the true origin of the change in $C_{para}$.

**C. TRANSITION MECHANISM OF VO$_2$**

We now proceed to examine the thermally triggered transition model. The charging time $t_c$ (and also the total heat accumulation time) will be the focus of the following discussions since it is closely related to the change of frequency. In contrast, the period of ($t_2 - t_c$) is dominated by the rate of heat dissipation and remains almost constant with respect to change in the bias current. The experimental relations between $t_c$ and the bias current for both Type A and Type B devices have been extracted from their respective oscillation waveforms and are shown as symbols in Fig. 6(a) – (c) with the device dimension indicated as figures beside the corresponding curves. A distinct feature for all devices is that the required heat accumulation time for triggering an IMT (i.e. $t_c$) decreases rapidly with the increase of the bias current. This behavior can be qualitatively understood as below. The heat produced by current-induced heating accumulates in the Pt/VO$_2$ bilayer during the period $0 < t < t_c$, and results in the increase of the device temperature. Once the critical transition temperature of the VO$_2$ is reached (t = $t_c$), IMT will be triggered. With a larger bias current, the Joule heating effect becomes more pronounced and the required heat accumulation time is shorter. It is assumed that a certain thickness of VO$_2$ ($t_{tran}$) directly underneath the Pt overlayer goes through transitions due to Joule heating in Pt and the IMT is corresponding to a critical unit-area heat accumulation ($Q_{area}$) in the VO$_2$. Based on the equivalent circuit in Fig. 4(b), the rate of heat generation in the DUT is determined by the current flowing in the Pt overlayer ($I_J$), and thus $Q_{area}$ can be written as:

$$Q_{area} = \int_0^{t_c} \alpha \frac{I_0^2 (1 - e^{-\frac{t}{\tau}})^2 R_{Pt}}{WL} dt + Q_{res}, \qquad (2)$$

where $\alpha \leq 1$ is a parameter to account for the heat dissipation through the sapphire substrate. The heat generated by the insulating VO$_2$ thin film is small and wide spread, and is thus not considered. Considering the gradual increase of the substrate temperature during heating, $\alpha$ is assumed to have the form of exp[-t/(c$\tau$)] where c is a unitless constant to indicate the different time scale of charging and heat accumulation processes. $Q_{res}$ is the residue heat from previous cycles of oscillations. Equ. (2) can be solved analytically to obtain the relationship between $I_0$ and $t_c$:

$$I_0 = \sqrt{\frac{QWL(2c+1)(c+1)}{R_{Pt}\, c\tau e^{-\frac{t_c}{c\tau}}[2c^2 e^{\frac{t_c}{c\tau}} - (c+1)e^{-\frac{2t_c}{\tau}} + 2(2c+1)e^{-\frac{t_c}{\tau}} - (2c+1)(c+1)]}}, \quad (3)$$

where Q is defined as $Q_{area} - Q_{res}$. Equation (3) is then used to fit the experimental $t_c - I_0$ relationship (symbols) in Fig. 6(a) – (c) with Q and c as the fitting parameters. The optimum fitting curves are shown as solid curves. Experimental values have been used for all other parameters during the fitting. It can be seen that Equ. (3) is able to reproduce the general trend of $t_c$ for both types of devices. The values of Q and c used for the fitting are shown in the respective insets. At this point of discussion, it is worth noting that Q is mainly determined by the relation $\Delta T \times C_{VO2} \times t_{tran} - Q_{res}$, where $\Delta T$ and $C_{VO2}$ are the change in temperature and heat capacity, respectively. Since both $t_{tran}$ and $Q_{res}$ can be different among devices with different dimensions and their values are unknown in the current experimental setup, Q can also vary among different devices. The relatively large fitted Q is presumably caused by the difference between the actual and the assumed time-dependence of α.

To investigate the possibility of electric field triggering $VO_2$ transition in our devices, similar electrical measurements have been performed on bare $VO_2$ samples in the same procedure and under the same conditions. It is found that bare $VO_2$ sample with L > 4 μm is unable to generate oscillation before discharge occurs inside $VO_2$ at a large bias current of over 10 mA. Figure 7(a) shows three cases of such discharge event with both probes moved aside (P1 and P2 show the probe positions). An interesting observation is that the discharge paths do not coincide with the expected electric field lines, suggesting the non-uniformity of the $VO_2$ thin film. When L is very small (~1.3 μm), electrical oscillations indeed have been observed on bare $VO_2$ and partially-Pt-covered samples [Fig. 7(b)]. One possible explanation is that leakage-current-induced heating brings the short-channel bare $VO_2$ sample above the onset of the thermally induced transition region.[35] The generation of oscillation then follows similar processes as described in the model in Section III(B). Nevertheless, further systematic studies are necessary to reveal the true generation mechanism of the oscillation observed in bare and partially-Pt-covered $VO_2$ samples with a small inter-probe distance and biased with a current source.

**IV. SUMMARY**

To conclude, we have fabricated Pt/$VO_2$ bilayer oscillators of different dimensions, in which the Pt overlayer serves the dual purposes of uniformly heating up the $VO_2$ and also weakening the electric field in (and voltage across) the $VO_2$. Reproducible electrical oscillations have been observed in systematic electrical measurements in an UHV environment. It is found that a higher oscillation frequency can be obtained with a larger bias current and/or smaller device dimensions. The charging duration after MIT is directly related to the oscillation frequency. A simple model based on current-induced heating in the Pt overlayer is proposed and is able to account for the experimental observations. Electrical measurements on bare $VO_2$ and partially-Pt-covered $VO_2$ performed under the same experimental conditions show that oscillation can only be triggered when the inter-probe distance is small, though the achievable frequency is lower than the case of Pt/$VO_2$ bilayer with a similar inter-probe distance. The results of this work help provide an alternative view in unraveling the $VO_2$ transition mechanism in $VO_2$-based oscillators.

**ACKNOWLEDGEMENTS**

YHW would like to acknowledge support by Ministry of Education, Singapore under its Tier 2 Grant (Grant No. MOE2013-T2-2-096).

**FIGURE CAPTIONS:**

FIG. 1. (Color online) (a) A schematic of the electrical measurements on a Pt/VO$_2$ bilayer oscillator. (b) and (c) SEM images of Type B (in dotted line) and Type A devices taken during the measurements, respectively. The scale bars in (b) and (c) are 100 μm and 10 μm, respectively.

FIG. 2. (Color online) (a) Discrete Fourier transform of the oscillation (inset) obtained by passing a DC current of 5.6 mA through a 2 μm × 40 μm device. The waveform around the onset of oscillation is shown in (b), in which all curves except the lowest one have been vertically shifted for the sake of clarity. The corresponding bias current is shown in legend in unit of mA.

FIG. 3. (Color online) (a) – (b) Dependence of the oscillation frequency on the bias current and channel length for Type A device. The dotted curves in (a) show the boundary of the oscillation window (Region II). (c) Dependence of the frequency on the bias current and channel width for Type B device. The oscillation window is shown as the shaded region (II) in the inset.

FIG. 4. (Color online) (a) A schematic of the thermally triggered oscillation in Pt/VO$_2$ bilayer. The simplified equivalent RC circuit is shown in (b). (c) and (d) show one cycle of the typical experimental oscillation waveforms (symbols) from Type A and Type B devices, respectively. All curves except for the lowest one have been shifted upwards for clarity. Solid curves are the fitting curves.

FIG. 5. (Color online) Dependence of the resistance of Pt strip (circle) and the time constant (triangle) on the dimensions of (a) Type A and (c) Type B devices. Solid curves are trend lines. The calculated parasitic capacitance is shown in (b) and (d).

FIG. 6. (Color online) Dependence of the experimental charging time (symbols) on both the bias current and the device dimensions for (a) Type A and (b) – (c) Type B devices. Solid curves are the fitting curves. The insets of (a) and (b) shows the respective values of Q and c used in the fitting.

FIG. 7. (Color online) (a) Typical SEM images of the VO$_2$ after breaking down under an intense electric field with both probes moved aside. The original probe positions are indicated by P1 and P2. The scale bars from left to right are 5, 5 and 10 μm. (b) Comparisons of the dependence of oscillation frequency on the bias current between bare VO$_2$ and Pt/VO$_2$ samples at an inter-probe distance of ~1.3 μm. Inset is a schematic of the measurement on partially-Pt-covered VO$_2$ samples.

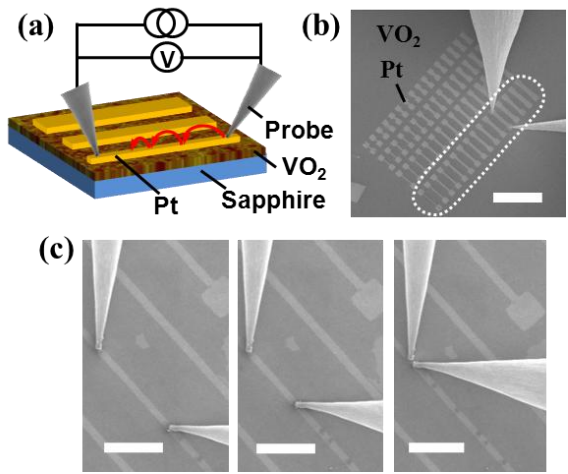

Fig. 1

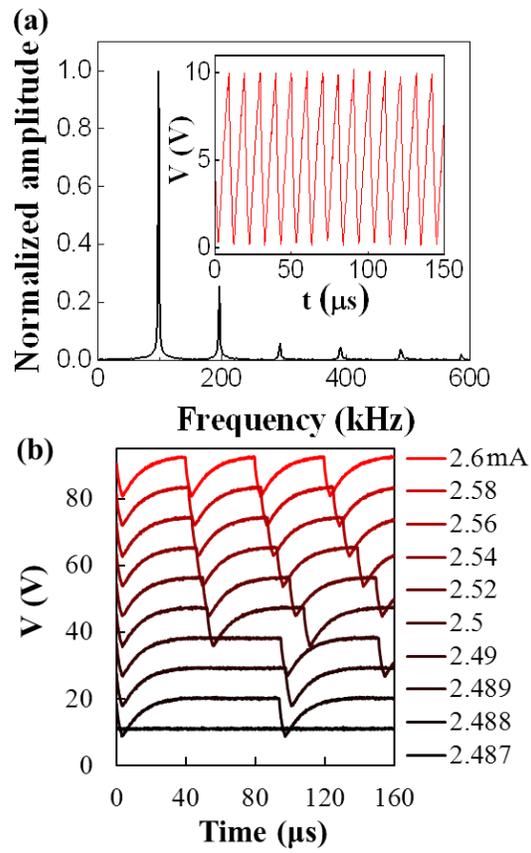

Fig. 2

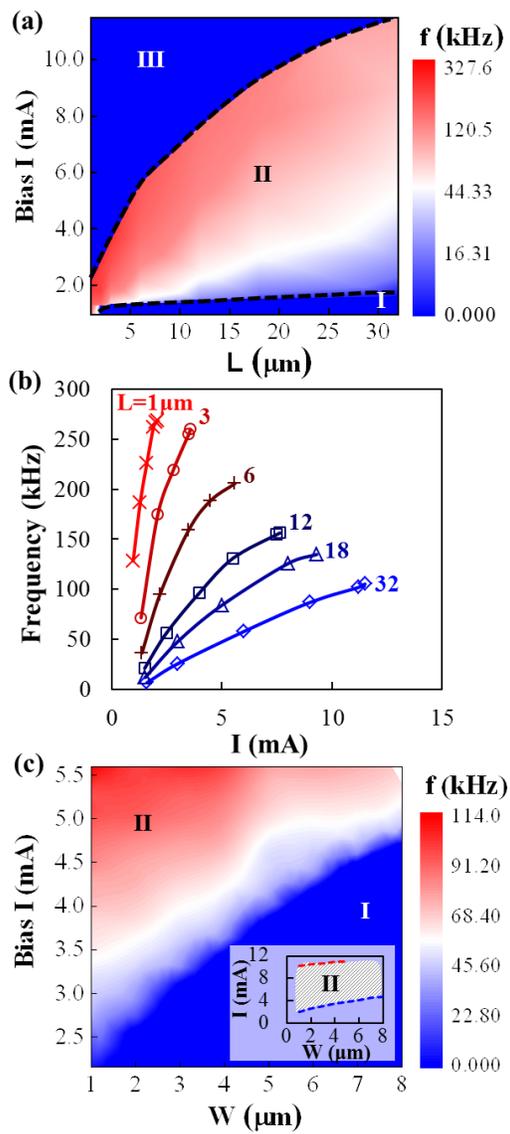

Fig. 3

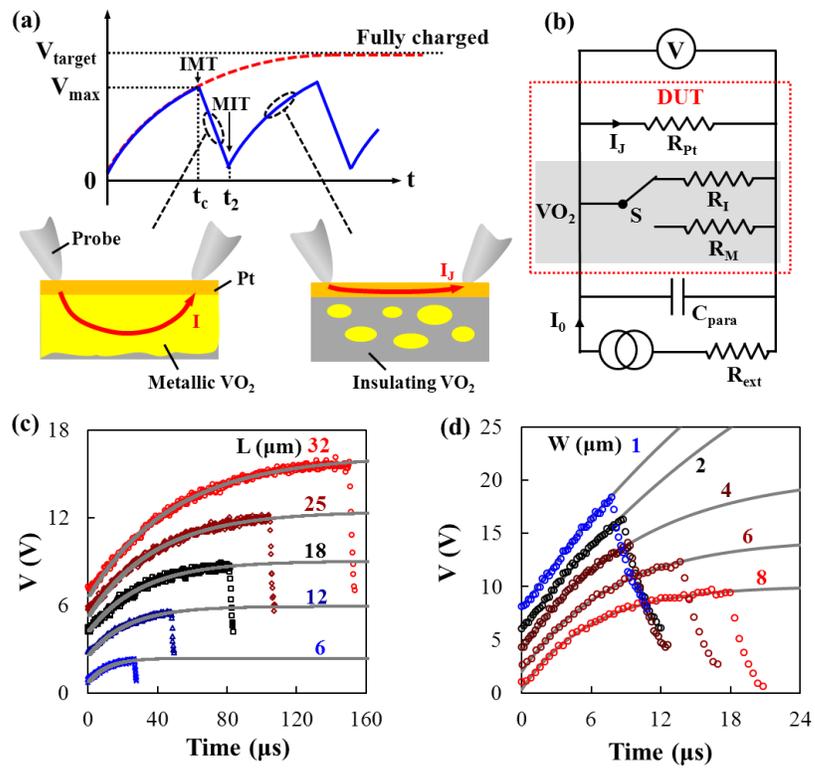

Fig. 4

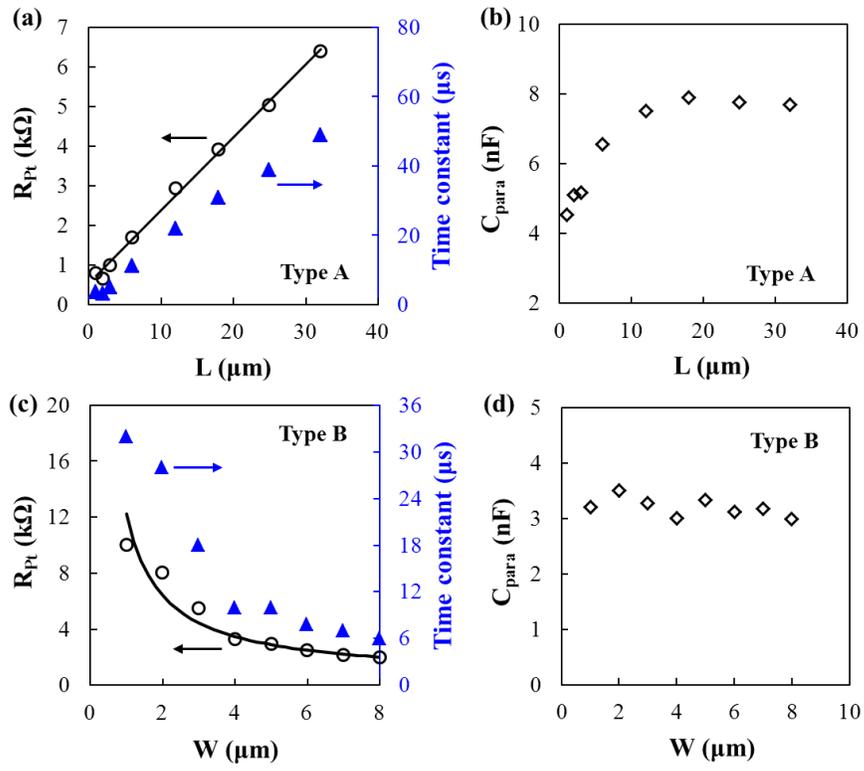

Fig. 5

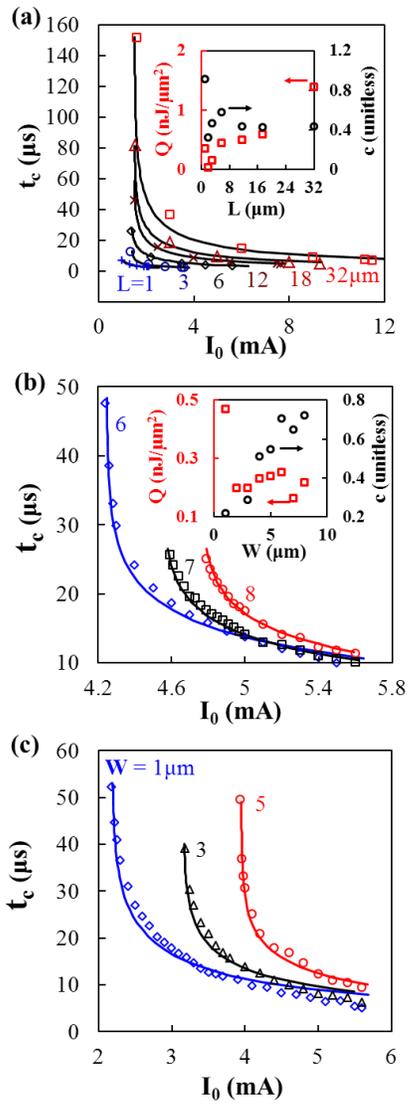

Fig. 6

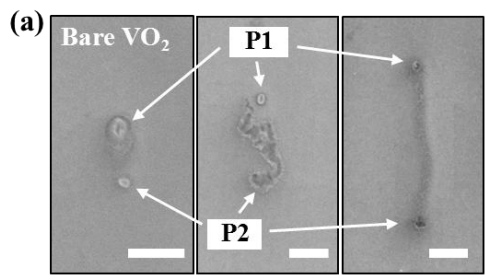

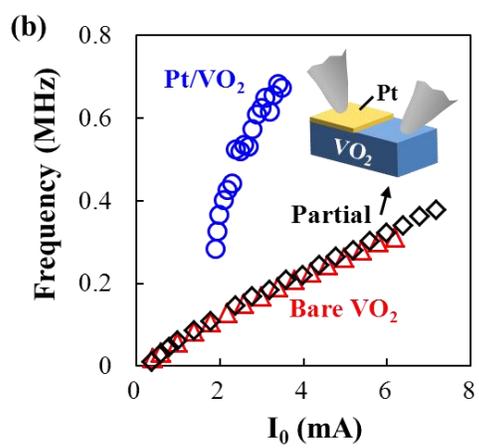

Fig. 7